\documentclass[12pt,a4paper,fleqn]{article}
\usepackage{epsf,amsfonts}
\usepackage[TS1,T1]{fontenc}
\usepackage[latin1]{inputenc}
\usepackage{mathrsfs}
\usepackage{amsmath}
\usepackage{amssymb}
\usepackage{epsfig}

\newcommand{\bg}{\begin{equation}}
\newcommand{\eg}{\end{equation}}

\def\d{\mbox{d}}
\def\Tr{\mbox{tr }} 
\def\m{_{\mu}}
\def\n{_{\nu}}
\def\M{^{\mu}}
\def\N{^{\nu}}
\def\mn{_{\mu\nu}}

\def\LD{{\cal L}}

\def\e{\mbox{e}}
\def\I{\mbox{i}}
\newcommand{\gl}[1]{(\ref{eq:#1})}
\newcommand{\Gl}[1]{Eq.~(\ref{eq:#1})}
\newcommand{\SL}[1]{\slash \hskip-.50em #1 \hskip+.05em}
\newcommand{\abs}[1]{\vert #1\vert}
\newcommand{\qut}[1]{``#1''}   
\newcommand{\Abb}[1]{Figure~\ref{fig:#1}} 
  
\begin{document}
  \begin{center}
    {\Large \textbf{Renormalization in Quantum Field Theory:\\[2pt] An Improved
        Rigorous Method}}\\[16pt] 
    Silke Falk$^*$, Rainer Häu\ss ling$^{**}$, and Florian Scheck$^{***}$\\ 
    \textit{Institut für Physik\\Johannes Gutenberg-Universität\\
    D-55099 Mainz (Germany)}\\
   (revised \today )
  \end{center}
  \vspace*{6pt}

  \begin{abstract}
    The perturbative construction of the $S$-matrix in the causal spacetime
    approach of Epstein and Glaser may be interpreted as a method of
    regularization for divergent Feynman diagrams. The results of any method of
    regularization must be equivalent to those obtained from the Epstein-Glaser
    (EG) construction, within the freedom left by the latter. In particular, the
    conceptually well-defined approach of Bogoliubov, Parasiuk, Hepp, and
    Zimmermann (BPHZ), though conceptually different from EG, meets this
    requirement. Based on this equivalence we propose a \textit{modified} BPHZ
    \textit{procedure} which provides a significant simplification of the
    techniques of perturbation theory, and which applies equally well to
    standard quantum field theory and to chiral theories. We illustrate the
    proposed method by a number of examples of various orders in perturbation
    theory. At the level of multi-loop diagrams we confirm that subdiagrams as
    classified by Zimmermann's forest formula in BPHZ can be restricted to
    subdiagrams in the sense of Epstein-Glaser, thus entailing an important
    reduction of actual computations. The relationship of our approach to the
    method of dimensional regularization (and renormalization) is particularly
    transparent, without having to invoke analytic continuation to unphysical
    spacetime dimension. It sheds new light on the role of some parameters that
    appear within dimensional regularization, and thus establishes a direct link
    of this traditional method to the BPHZ scheme.
  \end{abstract}

  \vfill
  $^{*}$ e-mail: falk@thep.physik.uni-mainz.de \\
  $^{**}$ e-mail: haeussli@thep.physik.uni-mainz.de \\
  $^{***}$ e-mail: scheck@thep.physik.uni-mainz.de
  \newpage

  \section{Introduction}
  As is well-known, formal perturbation theory applied to relativistic quantum
  field theory, in general, leads to ill-defined expressions for elements of the
  $S$-Matrix. Integrals that are expected to describe probability amplitudes for
  certain scattering processes are found to be divergent. Generally speaking,
  these ultra-violet (UV) divergencies can be traced back to the na\"{\i}ve
  application of time-ordering in the description of propagation of particles.
  There are two conceptually rather different lines of attack to deal with this
  problem: The first of these consists in a set of \textit{regularization}
  procedures all of which are designed such as to replace divergent integrals in
  Feynman diagrams by convergent ones in a consistent manner. These empirical
  regu\-larization schemes are justified by their usefulness in practical
  applications of quantum field theory to physical processes. In order to be
  consistent, they must fulfill all physical (normalization) conditions, order
  by order, or, at least, must contain enough freedom to meet these conditions.
  This is the essential prerequisite for the procedure of renormalization. In
  other terms, not every scheme of \textit{regularization} of divergent
  integrals of a given theory meets the stronger requirement of
  \textit{renormalizability} of that theory.

  The second line follows the approach developed by H.~Epstein and V.~Glaser,
  \cite{EG}, which is based on causality and locality in coordinate space. This
  procedure makes use of a well-defined rule for time ordering and thereby
  allows to construct an entirely divergence-free $S$-matrix from basic and
  general principles. The Epstein-Glaser (EG) approach is mathematically
  rigorous, within perturbation theory, but, when applied without modifications,
  is not very useful in practice. By its very construction, due to the process
  of distribution splitting, it contains a certain freedom which, subsequently,
  is fixed through its interpretation in terms of physics in the process of
  renormalization.  In fact, as was first proposed in \cite{Pinter,Prange}, the
  EG method can be interpreted itself as a regularization scheme. Thus, this
  approach is particularly useful as a reference framework for testing whether a
  given empirical method of regularization is physically admissible (in the
  sense of renormalizability), or not.
  
  Among the regularization procedures of the first group the classical method of
  Bogoliubov, Parasiuk, Hepp and Zimmermann (BPHZ), by its conceptual clarity,
  is the most rigorous one \cite{BoSh,Col}. The rules of regularization that it
  contains at the level of one-loop diagrams are equivalent to those of EG, with
  the latter suitably translated to momentum space. In addition BPHZ contains a
  general prescription, in the form of the \textit{forest formula}, for
  regularizing higher-loop diagrams. However, although its logical structure is
  transparent, the BPHZ procedure leads to rather involved integrals in explicit
  calculations which make it less suitable for practical computations of Feynman
  amplitudes as compared to more empirical methods such as dimensional
  regularization or the like. (In using the term ``dimensional regularization''
  we follow common conventions. In fact, this nomenclature means dimensional
  renormalization with, say, minimal subtraction.)
  
  With $d(x)$ a scalar distribution of singular order~$\omega$ the EG method
  defines advanced and retarded distributions through splitting of its support
  by a space-like hypersurface, say $v\cdot x=0$ with $v$ a timelike vector. As
  is well-known, a construction valid for all physically relevant values
  $\omega$ and for all test functions $g\in \mathfrak{S}^\prime (\mathbb{R}^k)$,
  is then
  \begin{subequations} 
    \begin{gather}
      \label{eq:1a}
      \int\!\!\d^4 x \; d_{\rm ret, reg}(x) g(x) 
      =  \int\!\!\d^4x\; d(x)\Theta (v\cdot x)\left( Wg\right) (x)\; , \\
      \label{eq:1b}
      \int\!\!\d^4 x \; d_{\rm adv, reg}(x) g(x) 
      =  -\int\!\!\d^4x\; d(x)\left[ 1-\Theta (v\cdot x)\right]\left( Wg\right)
      (x)\; ,
    \end{gather}
  \end{subequations}
  where the operator $W$ is defined through its action on test functions $g(x)$
  \begin{subequations}
  \begin{equation}
    \label{eq:2a}
    \left( Wg\right)(x) = g(x)-w(x)\sum_{\vert a\vert =0}^{\omega}
    \frac{x^a}{a!}\left( D^a g\right) (0)
  \end{equation}
  with $D^a$ the customary short-hand for partial derivatives 
  \begin{displaymath}
    D^a = \frac{\partial^{a_1+\ldots +a_k}}{\partial x_1^{a_1} \cdots \partial
      x_k^{a_k}}\; ,\quad \abs{a} = a_1+\ldots +a_k \; ,
  \end{displaymath}
  with $x^a$ the standard multicomponent notation for the coordinates, and with
  $w(x)$ a function satisfying the conditions
  \begin{equation}
    \label{eq:2b}
    w(0)=1\; ,\quad \left. D^a w\right|_{x=0} = 0\quad
    \mbox{for all} \quad 1\le \vert a\vert \le \omega \; .
  \end{equation}
  \end{subequations}
  A given choice of the function $w$ represents a specific regularization. This
  is the perspective adopted in~\cite{Pinter,Prange}. Any two different
  regularizations differ by a $\delta$-distribution and derivatives thereof up
  to the order $\omega$ in the integrands, \cite{SilkeDiss}, viz.
  \begin{gather*}
    \int\!\!\d^4 x \; \left( d_{{\rm reg,}w_1}(x)-d_{{\rm reg,}w_2}(x)\right)g(x)
    = \int\!\!\d^4 x \; \left( \sum_{\vert a\vert =0}^{\omega} c_a D^a \delta
      (x)\right)g(x)  \quad \mbox{with} \\
    c_a =  \int\!\!\d^4 x \; d(x) (-1)^{\abs{a}}\frac{x^a}{a!}\left[ w_2(x)-
      w_1(x)\right] \; .
  \end{gather*}
  This freedom of choice is essential for the subsequent renormalization process
  which relates the free parameters to values of physical parameters. As also
  shown in that work, a modified subtraction operator such as the one proposed
  in~\cite{JMGB} will yield a valid regularization but may turn out to be too
  restrictive for successful renormalization.
  
  In this paper we study a new method that we propose to call
  \textit{modified} BPHZ~\textit{procedure}. This method combines the practical
  usefulness of dimensional regularization with the structural simplicity of the
  classical BPHZ renormalization in the light of its equivalence to the
  Epstein-Glaser method. Like for any other empirical regularization method, the
  rigorous Epstein-Glaser framework is the landmark with respect to which the
  correctness and use of our modified procedure must be judged.
 
  The paper is organized as follows. In Sect.~2 we discuss the equivalence
  between the BPHZ and EG frameworks. In~Sect.~3 we describe the idea of the
  modified BPHZ~method and its justification, by means of its relationship to
  Epstein-Glaser. In Sect.~4 we give some instructive examples and work out the
  relationship to dimensional regularization. The final Sect.~5 gives a summary
  and outlook.
  
  \section{Equivalence of BPHZ and of EG frameworks}
  The BPHZ scheme is based on Feynman rules in momentum space. Schematically,
  and at this point still somewhat formally, a given diagram $\gamma$ with
  internal momenta $k$ is translated to an integrand of the form
  \begin{equation}
    \label{eq:3}
    I_\gamma (p,k) = \prod_{l\in\mathcal{L}} \Delta_c(p,k)\prod_{V\in\mathcal{V}}
    P_V(p,k)\; ,
  \end{equation}
  where $p$ denotes the set of external momenta, while $k$ stands for the
  internal momenta to be integrated over. The factors $\Delta_c$ are
  proportional to Feynman pro\-pa\-ga\-tors $\widetilde{\Delta}_F$ in momentum
  space, and correspond to the internal lines~$l$ of a given set~$\mathcal L$.
  The momentum flow is defined by the conventions chosen in the forest formula.
  A given vertex $V$ in the set $\mathcal V$ of vertices contributes the factor
  $P_V$. Consider an arbitrary irreducible one-loop diagram whose degree of
  divergence is $d(\gamma )$. The BPHZ approach replaces the integrand by the
  modified expression
  \begin{subequations}
  \begin{gather}
    \label{eq:4a}
    R_\gamma (p,k) = \left( 1 - t_p^{d(\gamma )} \right) I_\gamma (p,k)\; , \\
    \label{eq:4b}
    \mbox{where }t_p^{d(\gamma )} = \sum_{\abs{n}=0}^{d(\gamma )}
    \frac{1}{n!}p^n \left.\frac{\d}{\d p^n}\right|_{p=0} \; .
  \end{gather}
  \end{subequations}
  The Taylor operator $t_p^{d(\gamma )}$ stands symbolically for the expansion
  in terms of the set of independent external momenta $p$. In view of subsequent
  renormalization, the general result of regularization has the form
  \begin{equation}
    \label{eq:5}
    \left. \int\!\!\d^4k\; I_\gamma (p,k) \right|_{\rm BPHZ, reg}
    =  \int\!\!\d^4k\; R_\gamma (p,k) + P^{(d(\gamma ))}(p) \; ,
  \end{equation}
  with $P^{(d(\gamma ))}(p)$ a polynomial of degree $d(\gamma )$ representing
  the remaining freedom.
  
  For the sake of illustration we will refer repeatedly to $\phi^4$~theory in
  which case $P_V$ yields a power of the coupling constant $g$. As an example,
  consider the four-point function of $\phi^4$~theory at one loop i.e. the
  diagram shown in~\Abb{1}, with external momentum $p$ and containing two
  internal lines. In this case $d(\gamma )=0$ and BPHZ regularization yields 
  \begin{figure}[htbp]
  \begin{center}
    \epsfig{file=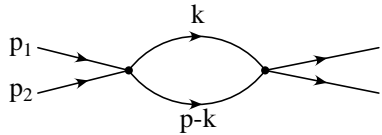,width=5cm} 
  \caption{Four-point function in the $\phi^4$~model, at one loop, with external
    momentum $p=p_1+p_2$} \label{fig:1}
  \end{center} 
  \end{figure}
  \begin{equation}
    \label{eq:6}
    \left. \widetilde{\Delta}^2_{\rm F}(p)\right|_{\rm BPHZ, reg}
    = -\frac{1}{(2\pi )^6}\int\!\!\d^4k\; \left\{ \frac{1}{k^2-m^2}
      \frac{1}{(p-k)^2-m^2} - \frac{1}{(k^2-m^2)^2} \right\} \; .
  \end{equation}
  In order to compare with the corresponding result of EG regularization the
  equations~\gl{1a} and \gl{2a}, as well as all test functions are transformed
  to momentum space, by Fourier transform, so as to obtain
  \begin{equation}
    \label{eq:7}
    d_{\rm reg}(g) = \int\!\!\d^4x\; d_{\rm reg}(x)g(x)
                   = \int\!\!\d^4k\; \widetilde{d}_{\rm reg}(k)\widetilde{g}(k)\; ,
  \end{equation}
  as well as the analogue of \gl{6},
  \begin{align}
    \left. \widetilde{\Delta}^2_{\rm F}(p)\right|_{\rm reg} & 
    = -\frac{1}{(2\pi )^6}\int\!\!\d^4k\; \frac{1}{k^2-m^2} \left\{ 
    \frac{1}{(p-k)^2-m^2} \right. \nonumber \\
    \label{eq:8}
    & - \left.\frac{1}{(2\pi )^2}\int\!\!\d^4p^\prime\;
      \frac{\widetilde{w}(p^\prime )}{(p'-k)^2-m^2} \right\} \; .
  \end{align}
  In the present example one may choose the function $w$ to be $w(x)=1$, as a
  limiting case. Obviously, it satisfies the conditions \gl{2b}. Furthermore,
  its Fourier transform being $\tilde{w}(p)=1/(2\pi )^2\delta (p)$, it is seen
  to yield the BPHZ expression \gl{6}.
  
  Of course, proving the equivalence of the BPHZ and EG schemes, beyond the
  one-loop level and for other theories, becomes technically more complicated.
  As is well-known, this is due to the fact that EG is an expansion in terms of
  the number $n$ of vertices, i.e. in terms of powers of the coupling constant,
  while BPHZ is an expansion in terms of loops, i.e. a formal expansion in terms
  of Planck's constant.
  \section{A modified BPHZ procedure}
  \label{sec:3}
  
  In the classical BPHZ method divergent momentum integrals are regularized by
  means of appropriate Taylor subtractions of the integrand. Although, on the
  basis of Zimmermann's forest formula, this approach is transparent and
  well-defined in principle, its practical implementation in higher orders of
  perturbation theory is cumbersome. From a practical point of view, other
  methods of regularization such as analytic continuation in the dimension of
  space-time are better tools in actual calculations.
  
  The alternative procedure that we propose aims at modifying the well-defined
  framework of BPHZ in such a way that it becomes as practicable as customary
  dimensional regularization. In essence, the idea is to introduce Feynman
  parameters at the level of the \textit{unsubtracted} integrand and to apply
  Taylor subtraction to the modified integrand only. As we will show this
  preserves the mathematical rigour of the BPHZ scheme but simplifies enormously
  subsequent integrations over internal momenta.
  
  To start with and in order to explain the essence of the modified method we
  give a very simple example from the $\phi^4$~model. Within the BPHZ scheme and
  at second order in the coupling constant $g$, the contribution of the one-loop
  diagram to the four-point function is logarithmically divergent. BPHZ
  regularize it by Taylor subtraction of the integrand to order zero, viz.
  \begin{equation}
    \label{eq:9}
    \frac{1}{2}g^2 \frac{1}{(2\pi )^4}\int\!\!\d^4k\; 
    \left( 1-t_p^0\right) \frac{1}{[k^2-m^2]}\frac{1}{[(p-k)^2-m^2]}
    =: \Lambda (p)
  \end{equation}
  and thus obtain a well-defined expression. In a first step, and in analogy to
  dimensional regularization, we parametrize the unmodified integrand by means
  of a Feynman parameter~$z$, such that the integrations over $z$ and over the
  internal momentum $k$ may be interchanged. In a second step the integration
  variable is subject to a translation by the vector $(z-1)p$, $k\mapsto
  q=k+(z-1)p$, so that mixed terms containing external and internal momenta no
  longer appear.  Finally, the Taylor subtraction is applied to the modified
  integrand. The three steps are given by, respectively,
  \begin{align}
    \label{eq:10}
    \Lambda (p) & = \frac{g^2}{2(2\pi )^4} \int\!\!\d^4k\; 
                    \left( 1-t_p^0\right)\int_0^1\!\!\! \d z\;  
                    \frac{1}{\left\{ [(p-k)^2-m^2](1-z) + z[k^2-m^2]\right\}^2 }
                    \nonumber \\
                & = \frac{g^2}{2(2\pi )^4}\int_0^1\!\!\! \d z\! \int\!\!\d^4k\; 
                    \left( 1-t_p^0\right)
                    \frac{1}{[k^2-2pk(1-z)+p^2(1-z)-m^2]^2} \nonumber \\
                & = \frac{g^2}{2(2\pi )^4}\int_0^1\!\!\! \d z\!
                    \int\!\!\d^4q\; \left( 1-t_p^0\right)
                    \frac{1}{[q^2+z(1-z)p^2 - m^2]^2} \; . 
  \end{align}
  Note that the scaling behaviour of the integrand for large values of~$k$
  remains unchanged by the introduction of a Feynman parameter. Therefore, the
  coefficients of the Taylor expansion of order higher than the degree of
  divergency lead to convergent integrals. The calculation of the integral over
  $q$ is standard. Making use of a Wick rotation, one obtains
  \begin{subequations}
  \begin{equation}
    \label{eq:11a}
    \Lambda (p) = \frac{\I g^2}{32\pi^2}\int_0^1\!\!\! \d z\;
    \ln\left(\frac{m^2}{m^2-z(1-z) p^2}\right) \; .
  \end{equation}
  Note that, unlike in dimensional regularization, the result \gl{11a} is
  exclusively obtained in dimension four of physical space-time.
  Furthermore, the remaining freedom in the approach discussed here which allows
  for a constant additive term (with respect to $p$), may be made explicit by
  replacing $m\mapsto \mu$, with $\mu$ an arbitrary mass, in the numerator
  of the logarithm in~\gl{11a}, i.e. $\Lambda (p)$ may be replaced by
  \begin{equation}
    \label{eq:11b}
    \Lambda^{(\mu )} (p) = \frac{\I g^2}{32\pi^2}\int_0^1\!\!\! \d z\;
    \ln\left(\frac{\mu^2}{m^2-z(1-z) p^2}\right) \; .
  \end{equation}
  Indeed, the expressions \gl{11a} and~\gl{11b} differ by a constant only.  For
  instance, the specific choice $\mu^2=4\pi \mu^2_{{\rm dim.reg}}\e^{-\gamma}$,
  with $\gamma$ Euler's constant, reproduces the well-known result of
  dimensional regularization, 
  \begin{equation}
    \label{eq:11c}
    \Lambda^{({\rm dim.reg})} (p) = \frac{\I g^2}{32\pi^2}\left\{
    -\gamma + \int_0^1\!\!\! \d z\; \ln\left(\frac{4\pi\mu^2_{{\rm
            dim.reg}}}{m^2-z(1-z) p^2}\right) \right\} \; .
  \end{equation}
  \end{subequations}

  Somewhat more generally, a convergent one-loop integral whose integrand was
  Taylor subtracted to the appropriate order $\omega$
  \begin{subequations}
  \begin{equation}
    \label{eq:12a}
    J_\gamma (p) = \int\!\!\d^4k\left( 1-t_p^\omega \right)I_\gamma (k,p) \; ,
  \end{equation}
  is transformed by a translation of the argument, $k\mapsto q=k+\lambda p$,
  \begin{equation}
    \label{eq:12b}
    J_\gamma (p)  = \int\!\!\d^4k\; \left\{ \left( 1-t_p^\omega \right)
    I_\gamma (k+\lambda p,p) - \Delta (q,p) \right\} \; ,   
  \end{equation}
  \end{subequations}
  where the function $\Delta (q,p)$ is defined by this equation and, hence, is
  given by
  \begin{subequations}
  \begin{equation}
    \label{eq:13a}
    \Delta (q=k+\lambda p,p) = \left( 1-t_p^\omega \right) 
                               I_\gamma (k+\lambda p,p)
    - \left.\left( 1-t_p^\omega \right) I_\gamma (q,p)\right|_{q=k+\lambda p}\; .
  \end{equation}
  Note that in the first term the Taylor operator $t_p^\omega$ applies to both
  arguments of the function $I_\gamma$, while in the second term it applies to
  the second argument only. 
  
  Alternatively, the function $\Delta (q,p)$, with $q=k+\lambda p$ can also be
  written as follows:
  \begin{equation}
    \label{eq:13b}
    \Delta (q=k+\lambda p,p) = \left( 1-t_p^\omega \right)
        \left[ \left.\left( t_p^\omega
        I_\gamma (q,p)\right)\right|_{q=k+\lambda p}\right] \; .
  \end{equation}
  \end{subequations}
  The result \gl{13b} follows from an identity for the Taylor operator
  $t_p^\omega$ in the variable $p$ about the point $p=0$, applied to a
  differentiable function $F$ of two variables,
  \begin{equation}
    \label{eq:14}
    t_p^\omega F(k+\lambda p,p) = t_p^\omega \left(
      \left. t_p^\omega F(q,p)\right|_{q=k+\lambda p} \right) \; .
  \end{equation}
  Indeed, denoting by $\partial_1$ and $\partial_2$ the derivatives with respect
  to the first and second argument of $F$, respectively, the left-hand side is
  \begin{displaymath}
    t_y^N F(x+\lambda y,y) = \sum_{i=0}^N \binom{i}{N} \lambda^iy^i 
      \left(\partial_1^i\partial_2^{N-i}F(x,0)\right) \; .
  \end{displaymath}
  The right-hand side, in turn, is computed to be
  \begin{gather*}
    t_y^N\left( \left. t_y^NF(u,y)\right|_{u=x+\lambda y} \right) 
    = t_y^N\left\{ \sum_{k=0}^N \frac{1}{k!} y^k \left( \partial_2^k
        F(u,y)\right)_{u=x+\lambda y,y=0} \right\} \\
    = \sum_{i=0}^N \binom{i}{N} \sum_{k=0}^N \frac{1}{k!}\lambda^i y^i
      \left( \partial_1^i\partial_2^k F(x,0)\right)
      \left.\left( \partial_y^{N-i}y^k\right)\right|_{y=0}
  \end{gather*}
  With $\left.(\partial_y^{N-i}y^k)\right|_{y=0}=k!\delta_{N-i,k}$ this is seen
  to be the same expression as above.

  As a result, the integral $J_\gamma$, after translation of the internal
  momentum, takes the form
  \begin{equation}
    \label{eq:15new}
    J_\gamma (p) = \int\!\!\d^4k\; \left( 1-t_p^\omega \right)
    I_\gamma (k+\lambda p,p) + J^{(0)}(p) \; ,
  \end{equation}
  where $J^{(0)}$ is the integral over $\Delta$. Closer examination of~\gl{13a}
  shows that this integrand can be written as a sum of derivatives with respect
  to $k$ of order one and higher, and, hence, gives rise to surface terms which
  vanish at infinity.  Thus, $J^{(0)}$ vanishes. This calculation demonstrates
  that translation of the integration variable is an admissible operation.
  
  As will be clear from the examples worked out below, the parameter $\lambda$,
  in general, is a function of the Feynman parameter(s) $z$.  Like in the
  example above the translation is chosen such that mixed terms in $k$ and $p$
  disappear. The integrand is then a function of $k^2$ only so that the
  integration can be done in Euclidean polar coordinates, via Wick rotation.

  These examples motivate the following modified BPHZ procedure:
  \begin{enumerate}
  \item In a given integral $I_\gamma (k,p)$ with external and internal momenta
    $p$ and $k$, respectively, introduce integral representations by means of a
    set of Feynman parameters $z$, interchange the $z$ integrations with the
    operator $( 1-t_p^\omega )$, and with the integration over the internal
    momenta $k$.
  \item Perform a translation of the $k$-variables such that internal and
    external momenta are decoupled. 
  \item To the integrand apply Taylor subtraction up to singular order $\omega$.
  \item Do the $k$-integrals by means of Wick rotation and using Euclidean polar
    coordinates.
  \item In order to make contact with dimensional regularization replace the
    mass parameter(s) by general constant(s) $\mu$ so that only additive terms
    appear which form a polynomial in~$p$ up to and including the singular
    order~$\omega$. In some cases this does not exhaust the freedom necessary
    for renormalization because, obviously, the modified BPHZ~method has the
    same number of parameters as the original one. This is essential for
    identifying the physical parameters of the theory (masses, charges etc) in
    each order of perturbation theory.
  \end{enumerate}
  This is a well-defined algorithm whose advantages are evident. The general
  analysis given in eqs.~\gl{12a}--\gl{15new} as well as the examples at second
  and higher orders, lend strong support to the conjecture that it meets all
  requirements of physical renormalization.  Its closeness to the original BPHZ
  regularization and, hence, to Epstein-Glaser regularization, guarantees that
  it is an admissible regularization scheme.
  
  By a suitable choice of the parameter(s) $\mu$ one makes contact with
  well-known regularization methods such as dimensional regularization, without
  having to continue to unphysical space-time dimensions. The method is
  mathematically rigorous but more practicable than the original BPHZ approach.
  
  Furthermore, turning to fermions, no continuation of the Clifford algebra of
  Dirac $\gamma$-matrices is necessary given the fact that the modified BPHZ
  method works exclusively in dimension four.
  
  Our approach is rather close to the framework of Epstein and Glaser, but
  allows for direct comparison with unmodified BPHZ. Due to cancellations of a
  certain class of subdiagrams there are important simplifications in the
  calculation of higher-order processes. In order to explain this point we need
  some preparation and definitions.
  
  As we stated above, EG is an expansion in terms of the coupling constant $g$,
  while BPHZ is an expansion in powers of $\hbar$, hence in terms of the number
  of loops. EG constructs a functional $T_n$ describing a diagram with $n$
  vertices by recurrence from the tempered distribution $T_1=\I\LD_{\rm int}$.
  The total diagram depends on functionals which were regularized previously at
  orders lower than $n$, say $m<n$. Thus, the corresponding subdiagrams contain
  irreducible parts with a number of vertices \textit{smaller} than $n$. We
  shall call such subdiagrams \textit{Epstein-Glaser-subdiagrams} or, for
  short, EG~subdiagrams. The BPHZ framework, in turn, works by successive
  addition of counter terms proportional to ascending powers of~$\hbar$ and, as
  a consequence, requires a different classification of subdiagrams. Let us
  call \textit{BPHZ~subdiagram} any irreducible divergent part of the total
  diagram which contains a smaller number of \textit{loops} than the main
  diagram. In particular, there will be sub\-diagrams which are lower in loop
  order but do not have a smaller number of vertices. We call these
  \textit{pure} BPHZ~subdiagrams. An example we shall study in more detail
  below is the \qut{sunrise} diagram in the $\phi^4$~model, cf.~\Abb{2}. In the
  framework of BPHZ it contains three logarithmically divergent subdiagrams. In
  the perspective of EG, in contrast, it is a diagram with two vertices and,
  hence, contains no divergent subdiagram at all. In our terminology the three
  BPHZ~subdiagrams are \textit{pure} BPHZ~subdiagrams. The sum of the counter
  terms generated by these subdiagrams does not contribute to the regularization
  of the sunrise diagram. This example as well as other examples studied
  in~\cite{SilkeDiss} confirm this to be a general rule, and are in accordance
  with a theorem by Zimmermann~\cite{Zimmer}: Pure
  BPHZ~subdiagrams do not yield counter terms, i.e. their sum vanishes, and,
  thus, they may be left out in the modified approach.
  \begin{figure}[htbp]
  \begin{center}
    \epsfig{file=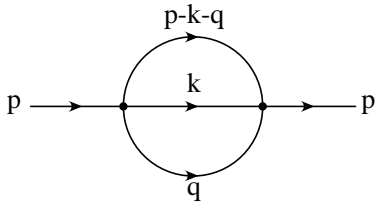,width=5cm} 
  \caption{Sunrise diagram in the $\phi^4$~model} \label{fig:2}
  \end{center} 
  \end{figure}
  
  We illustrate the method by a number of significant examples in second and
  higher orders.
  \section{Examples}
  \label{sec:4}
  We start with some classical examples from quantum electrodynamics and
  electroweak interactions, the self-energy of the electron, the vacuum
  polarization, and the vertex correction at one-loop order, then mention
  briefly the case of the triangle anomaly. We finish with a typical
  second-order, two-loop, process and with some remarks about higher-order
  processes which illustrate the simplicity of our alternative scheme. In all
  these examples the equivalence to EG regularization proves the correctness of
  the modified BPHZ approach.
  \subsection{Quantum electrodynamics with electrons}
  \label{sec:4.1}
  In the original BPHZ framework the \textit{self-energy} of the electron reads
  \begin{subequations}
  \begin{equation}
    \label{eq:15a}
    \Sigma (p) = -\frac{\I e^2}{(2\pi )^4} \int\!\!\d^4k\; 
                 \left( 1-t_p^1\right) \frac{\gamma\m (\SL{p}-\SL{k}+m)\gamma\M}
                 {[(p-k)^2-m^2]k^2} \; .
  \end{equation}
  In the modified BPHZ approach we introduce a Feynman parameter $z$,
  interchange integrations, and substitute $k\mapsto q=k-zp$ such as to decouple
  internal and external momenta, to obtain
  \begin{equation}
    \label{eq:15b}
    \Sigma (p) = -\frac{\I e^2}{(2\pi )^4} \int_0^1\!\!\!\d z \int\!\!\d^4 q\;
     \left( 1-t_p^1\right) \frac{\gamma\m ((1-z)\SL{p}-\SL{q}+m)\gamma\M}
     {[q^2-zm^2+z(1-z)p^2]^2} \; .
  \end{equation}
  This is easily worked out to be
  \begin{equation}
    \label{eq:15c}
    \Sigma (p) = \frac{e^2}{16\pi^2} \int_0^1\!\!\!\d z \;
    \left[ (z-1)2\SL{p}+4m\right]\, \ln\left(\frac{m^2}{m^2-(1-z)p^2}\right)\; .
  \end{equation}
  The remaining freedom is made explicit by replacing $m^2$ by an arbitrary
  squared mass $\mu^2$, 
  \begin{equation}
    \label{eq:15d}
    \Sigma^{(\mu )} (p) = \frac{e^2}{16\pi^2} \int_0^1\!\!\!\d z \;
    \left[ (z-1)2\SL{p}+4m\right]\, \ln\left(\frac{\mu^2}{m^2-(1-z)p^2}\right)\; .
  \end{equation}
  \end{subequations}
  One verifies that the choice
  \begin{equation}
    \label{eq:16}
    \mu^2=4\pi\mu^2_{\rm dim.reg}\, \e^{(1/2-\gamma )}
  \end{equation}
  reproduces (the finite part of) the result known from dimensional
  regularization, see e.g.~\cite{Ryder}.

  Lowest order \textit{vacuum polarization} in original BPHZ is given by
  \begin{subequations}
    \begin{equation}
      \label{eq:17a}
      \Pi\mn (p) = \frac{\I e^2}{(4\pi )^2}\int\!\!\d^4 q\; \left(
        1-t_p^2\right) \Tr\left(\gamma\m\frac{\SL{q}+m}{q^2-m^2}
        \gamma\n\frac{\SL{q}-\SL{p}+m}{(q-p)^2-m^2}\right)
    \end{equation}
    In the modified scheme, we introduce a Feynman parameter $z$, interchange
    the integration over~$z$ with the one over the internal momentum $q$, and
    perform a translation of the integration variable $q\mapsto
    \bar{q}=q-(1-z)p$, to obtain
    \begin{equation}
      \label{eq:17b}
      \Pi\mn^{(\mu )} (p) = -\frac{e^2}{2\pi^2}\int_0^1\!\!\!\d z\;
      \left( g\mn p^2-p\m p\n\right) z(1-z)
      \ln\left(\frac{\mu^2}{m^2-z(1-z)p^2}\right) \; .
    \end{equation}
  \end{subequations}
  As before, in order to exhaust the remaining freedom, we have replaced the
  numerator $m^2$ in the logarithm by an arbitrary squared mass $\mu^2$. The
  (finite part of) the known result of dimensional regularization~\cite{Ryder}
  is recovered by the choice
  \begin{equation}
    \label{eq:18}
    \mu^2=4\pi\mu_{\rm dim.reg}^2\, \e^{-\gamma}
  \end{equation}

  The \textit{vertex correction,} at the same order, finally, is found to be
  \begin{gather}
    \label{eq:19}
    -\I e\Lambda_\alpha^{(\mu )}(p,p')  = -\frac{\I e^3}{8\pi^2}\gamma_{\alpha}
    \int_0^1\!\!\!\d x\int_0^{1-x}\!\!\!\d y\; \ln\left(\frac{\mu^2(x+y)}{D^2}\right)
    + \frac{\I e^3}{8\pi^2}\gamma_{\alpha} \nonumber \\
     +  \frac{\I e^3}{16\pi^2} \int_0^1\!\!\!\d x\int_0^{1-x}\!\!\!\!\!\d y\;
    \frac{1}{D^2} \left\{ \gamma\n [(1-y)\SL{p}'-x\SL{p}+m]\gamma_{\alpha} [(1-x)\SL{p}
      -y\SL{p}' +m]\gamma\N\right\} \; ,
  \end{gather}
  where the denominator in the integrands stands for
  \begin{displaymath}
    D^2 = (x+y)m^2-x(1-x)p^2-y(1-y)p^{\prime\, 2} + 2pp'xy\; .
  \end{displaymath}
  As before, we replaced the numerator $m^2$ in the logarithm by an arbitrary
  term $\mu^2$, to cope with the remaining freedom after regularization. The
  analogous result in dimensional regularization is recovered by the same
  choice~\gl{16} as for the self-energy. This shows that the modified BPHZ
  regularization fulfills the Ward-Takahashi identity
  \begin{equation}
    \label{eq:20}
    \frac{\partial}{\partial p^\alpha} \Sigma (p) = - \Lambda_\alpha (p,p) \; .
  \end{equation}
  Though not surprising, this is a consistency check.
  \subsection{Chiral anomaly}
  \label{sec:4.2}
  \begin{figure}[htbp]
  We also analyzed the well-known vector-vector-axial vector (VVA) chiral
  anomaly sketched in~\Abb{3}, within the modified BPHZ procedure. 
   \begin{center}
    \epsfig{file=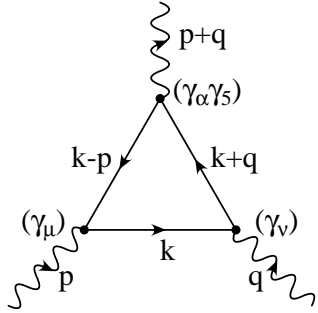,width=5cm} 
  \caption{Triangle graph contributing to the anomaly} \label{fig:3}
  \end{center} 
  \end{figure}
  Denoting the amplitude by $T_{\alpha\mu\nu}$ and choosing the internal loop
  momenta as shown in~\Abb{3}, conservation of the vector current at the two
  lower vertices should yield the Ward identities
  \begin{subequations}
  \begin{equation}
    \label{eq:22a}
    p^\mu T_{\alpha\mu\nu} = 0 \; ,\quad q^\nu T_{\alpha\mu\nu} = 0 \; ,
  \end{equation}
  whereas the axial current vertex should produce an anomalous Ward identity
  which survives even in the limit of the fermion mass~$m$ going to zero, viz.
  \begin{equation}
    \label{eq:22b}
   (p+q)^\alpha T_{\alpha\mu\nu} = 2mT_{\mu\nu} 
     + \frac{1}{2\pi^2} \varepsilon_{\mu\nu\sigma\tau} q^\sigma p^\tau \; ,
  \end{equation}
  \end{subequations}
  the term $T_{\mu\nu}$ being given by
  \begin{equation*}
    T_{\mu\nu} = \int\!\!\frac{\mbox{d}^4k}{(2\pi)^4}\, \mbox{tr} \left(
    \frac{\SL{k}+m}{k^2-m^2}\gamma_\mu \frac{\SL{k}-\SL{p}+m}{(k-p)^2-m^2}\gamma_5
    \frac{\SL{k}+\SL{q}+m}{(k+q)^2-m^2}\gamma_\nu\right)
    + \left( p\leftrightarrow q,\mu \leftrightarrow \nu \right) \, .
  \end{equation*}
  It is known that the anomaly can be shifted from the axial vector current to
  the vector current, or to a linear combination of these~\cite{WBar},
  \cite{Nov}. Thus, by requiring that it be the vector current which is
  conserved, some of the freedom in the renormalization process is made use of.

  The diagram of~\Abb{3} is linearly divergent. If regularized by Taylor
  subtraction, in the spirit of BPHZ, it is given by
  \begin{gather*}
    \tfrac{1}{2}T_{\alpha\mu\nu} =
    -\int\!\!\frac{\mbox{d}^4k}{(2\pi)^4}\,(1-t^1_{p,q})  \mbox{tr} \left(
    \frac{\SL{k}+m}{k^2-m^2}\gamma_\mu \frac{\SL{k}-\SL{p}+m}{(k-p)^2-m^2}
    \gamma_\alpha\gamma_5 
    \frac{\SL{k}+\SL{q}+m}{(k+q)^2-m^2} \gamma_\nu \right) \\ 
    = -2\int_0^1\!\!\!\mbox{d}x\int_0^{1-x}\!\!\!\mbox{d}y
        \int\!\!\frac{\mbox{d}^4k}{(2\pi)^4}\, (1-t^1_{p,q})  \\
    \frac{\mbox{tr}\left[ (\SL{k}+m)\gamma_\mu (\SL{k}-\SL{p}+m)
    \gamma_\alpha\gamma_5(\SL{k}+\SL{q}+m)\gamma_\nu\right]}
    {\left[ (k^2-m^2)(1-x-y) +((k-p)^2-m^2)x+((k+q)^2-m^2)y \right]^3} \\
    = -2\int_0^1\!\!\!\mbox{d}x\int_0^{1-x}\!\!\!\mbox{d}y
        \int\!\!\frac{\mbox{d}^4k}{(2\pi)^4}\, (1-t^1_{p,q})  \\
    \frac{\mbox{tr}\left[ (\SL{k}+m)\gamma_\mu (\SL{k}-\SL{p}+m)
    \gamma_\alpha\gamma_5(\SL{k}+\SL{q}+m)\gamma_\nu\right]}
      {\left[
          (k+(qy-px))^2-(qy-px)^2-m^2(1-x-y)+(q^2-m^2)y+(p^2-m^2)x\right]^3}\; .
  \end{gather*}
  A second term contributing to the chiral anomaly is obtained by interchanging
  $(p\leftrightarrow q), (\mu\leftrightarrow\nu )$. By the symmetry of the
  integrands this second term yields the same result as the first so that the
  factor $1/2$ on the left-hand side can be dropped.

  Following the rules of the modified BPHZ scheme described in Sect.~\ref{sec:3}
  one performs the substitution
  \begin{equation*}
    \bar{k} = k - (qy-px)
  \end{equation*}
  such as to separate internal and external momenta, and to allow for separation
  of terms even and odd in the new integration variable~$\bar k$. Indeed, only
  the even terms contribute to the integral. A straightforward calculation leads
  to the result
  \begin{equation*}
    T_{\alpha\mu\nu} = T_{\alpha\mu\nu}^{\rm log} 
                     + T_{\alpha\mu\nu}^{\rm finite}\; , 
  \end{equation*}
  the logarithmically divergent term and the finite term being given by,
  respectively,
  \begin{align*}
    T_{\alpha\mu\nu}^{\rm log} 
    & = \frac{1}{2\pi^2}\int_0^1\!\!\!\mbox{d}x\int_0^{1-x}\!\!\!\mbox{d}y\,
    \varepsilon_{\alpha\mu\nu\sigma} 
       \left\{ (3x-1)p^\sigma - (3y-1)q^\sigma\right\} \\
     & \ln\left( \frac{m^2}{m^2+(qy-px)^2-q^2y-p^2x} \right) \; , \\
    T_{\alpha\mu\nu}^{\rm finite}
    & = \frac{1}{2\pi^2}\int_0^1\!\!\!\mbox{d}x\int_0^{1-x}\!\!\!\mbox{d}y\,
    \Biggl\{\varepsilon_{\alpha\mu\nu\sigma} ((y-1)q^\sigma -
      (x-1)p^\sigma ) \\ 
    & + \frac{\varepsilon_{\alpha\mu\nu\sigma}\left\{ \left[
        (y-1)q^\sigma-(x-1)p^\sigma\right][(qy-px)^2-m^2] -yq^2p^\sigma +
      xp^2q^\sigma\right\} } 
        {m^2+(qy-px)^2-q^2y-p^2x} \\
    & + 2\frac{y\,\varepsilon_{\alpha\mu\sigma\tau}p^\sigma q^\tau
      [(y-1)q_\nu-xp_\nu ] + x \,\varepsilon_{\alpha\nu\sigma\tau} 
        q^\sigma p^\tau [ (x-1)p_\mu - yq_\mu] }
        { m^2+(qy-px)^2-q^2y-p^2x} \Biggr\} \; .
  \end{align*}
  In this example the well-known result obtained from unmodified BPHZ, or from
  dimensional regularization~\cite{Nov}, is obtained in a straightforward and
  technically simpler fashion. Replacing $m^2$ in the logarithmic integrand by
  an arbitrary parameter $\mu^2$ does not change the total expression. Indeed,
  this replacement produces an additive term proportional to
  \begin{displaymath}
    \left\{ (3x-1)p^\alpha - (3y-1)q^\alpha \right\}
    \ln\left(\frac{m^2}{\mu^2}\right)\; ,
  \end{displaymath}
  which yields zero after integrating over the Feynman parameter $y$ from~$0$ to
  $(1-x)$, and over $x$ from~$0$ to~$1$.

  For the sake of completeness we verify the Ward identities \gl{22a} and
  calculate the anomaly \gl{22b}. Straightforward calculation of the divergence
  $p^\mu T_{\alpha\mu\nu}$ leads to the result
  \begin{gather*}
    p^\mu T_{\alpha\mu\nu} = \frac{1}{2\pi^2}\varepsilon_{\alpha\mu\nu\sigma}
    p^\mu q^\sigma\int_0^1\!\!\!\mbox{d}x\int_0^{1-x}\!\!\!\mbox{d}y\, 
    \frac{1}{m^2+(qy-px)^2-q^2y-p^2x} \\
    \bigl[ (-yx+2yx^2+x^3-\tfrac{1}{2}x^2)p^2
    + (yx^2-y^2x)qp + (-y^3-2y^2x+\tfrac{1}{2}y^2+yx)q^2\bigr] \; .
  \end{gather*}
  In the diagram of~\Abb{3} the vector bosons at the lower vertices are
  identical so that $p^2=q^2$ (and equal to zero in the case of external
  photons).  With $q^2=p^2$ the integrand is antisymmetric under exchange of~$x$
  and~$y$ while the domain of integration is symmetric. Therefore, the integral
  vanishes and the first of the Ward identities \gl{22a} holds true. The second
  Ward identity follows from the first by the symmetry $(p\leftrightarrow q)$,
  $(\mu\leftrightarrow\nu )$.

  Regarding the divergence \gl{22b} which contains the anomaly, we find for the
  first term on the right-hand side
  \begin{equation*}
    2mT_{\mu\nu}(m) = -\frac{1}{\pi^2} \varepsilon_{\mu\nu\sigma\tau}
    q^\sigma p^\tau \int_0^1\!\!\!\mbox{d}x\int_0^{1-x}\!\!\!\mbox{d}y\,
    \frac{m^2}{m^2+(qy-px)^2-q^2y-p^2x} \; .
  \end{equation*}
  The left-hand side of~\gl{22b} is calculated along the lines of the
  procedure described above. We find the result
  \begin{align*}
    (p+q)^\alpha T_{\alpha\mu\nu} = & 
    -\frac{1}{\pi^2} \varepsilon_{\mu\nu\sigma\tau}
    q^\sigma p^\tau \\
    & \int_0^1\!\!\!\mbox{d}x\int_0^{1-x}\!\!\!\mbox{d}y\,
    \Bigl\{ \frac{m^2}{m^2+(qy-px)^2-q^2y-p^2x} - 1 \Bigr\} \; ,
  \end{align*}
  and, upon comparison with the previous formula,
  \begin{equation*}
    (p+q)^\alpha T_{\alpha\mu\nu} = 2mT_{\mu\nu}(m) 
      + \frac{1}{2\pi^2}\varepsilon_{\mu\nu\sigma\tau}q^\sigma p^\tau \; ,
  \end{equation*}
  which is, indeed, the anomalous identity~\gl{22b}.
  
  This example illustrates well the advantage of the modified BPHZ procedure, as
  compared to original BPHZ renormalization or to dimensional renormalization,
  by its simplification of the momentum integral. Furthermore, the equivalence
  to Epstein-Glaser regularization puts the modified procedure on solid
  ground. As compared to dimensional regularization, in particular, there is no
  need to introduce an analytic continuation of $\gamma_5$ to any other
  space-time dimension than four.

  We note in passing that the chiral limit~$m\to 0$, like in the usual BPHZ
  framework, requires a separate discussion. We do not treat this case in the
  present work. 
  \subsection{ Higher loop diagrams }
  \label{sec:4.3}
  The \qut{sunrise} diagram in the $\phi^4$~model, cf.~Fig.~2, provides an
  instructive example for the comparison of BPHZ and EG regularizations. Being a
  diagram with two vertices it contains no EG~subdiagrams at all. In the
  perspective of BPHZ, however, it contains three logarithmically divergent pure
  BPHZ subdiagrams. Thus, in the former case it is regularized in a single step
  by Taylor subtraction with respect to the external momentum, while in the
  latter, one would have to invoke the forest formula for identifying the
  counter terms stemming from the three divergent subdiagrams. This is to say
  that the modified approach which, in essence, is a practicable version of EG,
  is technically simpler, and, furthermore it uses the fact that, in accord with
  Zimmermann's theorem~\cite{Zimmer}, the contributions from all pure BPHZ
  subprocesses cancel.

  Regularizing the quadratically divergent diagram of~Fig.~2 by Taylor
  subtraction of the integrand one has
  \begin{subequations}
  \begin{equation}
    \label{eq:23a}
    \Sigma (p) = \frac{g^2}{6(2\pi )^8} \int\!\!\d^4q\int\!\!\d^4k\;
    \left( 1-t_p^2\right) \frac{1}{[(p-k-q)^2-m^2]}\,
    \frac{1}{k^2-m^2}\, \frac{1}{q^2-m^2} \; .
  \end{equation}
  One succesively introduces Feynman parameters for the internal momenta $k$,
  $q$, and $p-k-q$, and applies the necessary translations which decouple
  internal and external momenta. Details of this calculation are given in the
  appendix. The result is
  \begin{align}
    \Sigma (p) & = \frac{g^2}{6(4\pi )^4} \int_0^1\!\!\!\d z \int_0^1\!\!\!\d x\;
    \frac{(1-2z)(1-2x)p^2}{(z-1)(1-z+z^2)} \nonumber \\ 
    \label{eq:23b}
     & \ln\left(
    \frac{[z(1-z)(1-x)+x]m^2}{-xz(1-z)(1-x)p^2+z(1-z)(1-x)m^2+xm^2}\right) \; .
  \end{align}
  As in the previous examples one replaces the parameter $m^2$ by an arbitrary
  parameter $\mu^2$ but verifies that the result~\gl{23b} remains unchanged,
  \begin{displaymath}
    \Sigma^{(\mu )} (p) = \Sigma (p) \; .
  \end{displaymath}
  It is instructive to compare the result~\gl{23b} to a calculation of the
  sunrise diagram using dimensional regularization~\cite{vanHees}. The result is
  \begin{gather}
    \Sigma_{\rm dim.reg} (p)  = \frac{g^2}{6(4\pi )^4} \int_0^1\!\!\!\d z
    \int_0^1\!\!\!\d x\; \left\{ \left( -\frac{(1-x)}{x} 3m^2 + (1-x)p^2\right)\right.
    \nonumber \\
    \label{eq:23c}
        \left. \ln\left(
        \frac{[z(1-z)(1-x)+x]m^2}{-xz(1-z)(1-x)p^2+z(1-z)(1-x)m^2+xm^2}\right)  
        + \frac{1}{2}p^2 \right\} \; . 
  \end{gather}
  \end{subequations}
  The expressions \gl{23b} and~\gl{23c} both are regularizations of the same
  scalar distribution. Furthermore, their Taylor expansion vanishes up to the
  order $p^2$, in the first case by construction, in the second case due to the
  additional term $p^2/2$. As this exhausts the remaining freedom in
  regularizing, one concludes that the two results are identical.
  
  In order to make contact with the unmodified BPHZ procedure we have checked by
  explicit calculation using the forest formula that, indeed, the three
  subdiagrams which are not EG~subdiagrams cancel in the Taylor expansion and,
  hence, do not contribute to the regularization of the sunrise diagram. Thus,
  the modified BPHZ procedure is very close to pure Epstein-Glaser
  regularization and avoids from the start irrelevant contributions from
  pure BPHZ~subdiagrams to the regularized amplitudes, in agreement with the
  proof by Zimmermann \cite{Zimmer}.
  
  These results are corroborated by case studies of EG regularization in higher
  orders. Among others we studied the four-point function of the $\phi^4$~theory
  in dimension~$4$, at the level of two loops. The same model in dimension~$6$
  provides an example which besides yielding divergent EG~subdiagrams, also
  exhibits pure BPHZ~subdiagrams. The contribution of the pure BPHZ~subdiagrams,
  i.e. those which have no counterpart in EG, upon Taylor subtraction, are found
  to vanish, as expected. In the latter example we also studied three-loop
  contributions to the two-point function~\cite{SilkeDiss}.  In all cases
  EG~regu\-larization on one hand, and calculation following the forest formula
  restricted to EG~subdiagrams, on the other, yield identical results.
  \section{Conclusions and outlook}
  \label{sec:5}
  
  The modified BPHZ procedure that we advocate in this paper combines the
  transparent concept of BPHZ regularization with the practical usefulness of
  dimensional regularization. In particular, the relevant integrals are easier
  to calculate than the corresponding ones within the original BPHZ~method.
  Furthermore, the momentum dependent logarithms always contain a reference mass
  which is identical with the typical mass parameter of the theory (the electron
  mass in the case of the examples from QED, the scalar mass in the
  $\phi^4$~model). We showed, however, that rescaling is possible within the
  freedom allowed by the regularization process. In the examples with one loop,
  for instance, this allows to introduce a new mass parameter which may be
  identified with the parameter of dimensional regularization. However, there is
  an essential difference here: While dimensional regularization requires the
  introduction of this parameter for (spacetime-)dimensional reasons, in our
  approach it is a manisfestation of the general freedom within the process of
  regularization.  This remark, in turn, justifies its appearence in the results
  of dimensional regularization.
  
  The comparison of BPHZ regularization along the forest formula with the
  Epstein-Glaser construction confirms the expected significant simplification
  of explicit calculations in higher orders. In the light of different
  classifications of subdiagrams in the framework of BPHZ on one side, and in the
  Epstein-Glaser construction on the other, the summation over the subdiagrams
  contained in the forest formula is restricted to subdiagrams in the sense of
  Epstein-Glaser.\footnote{Of course, a certain choice of the standard momentum flow in
  the forest formula had to be made but the conclusion should be independent of
  that choice.} The modified procedure implies, in particular, that the
  combinatorics of higher-order diagrams is described by the restricted forest
  formula which takes account exclusively of the class of EG~subdiagrams. Thus,
  this method is as straightforward as, e.g. dimensional regularization, and has
  the virtue to rest on solid mathematical ground.
  \newpage
  \appendix
  \section{Derivation of \Gl{23b}}
  \label{sec:a1}

  The strategy for deriving~\gl{23b} goes as follows. A first Feynman parameter
  denoted by $z$ is introduced for the $k$-integration. A subsequent translation
  by $k\mapsto \tilde{k}=k-(1-z)(p-q)$ then frees this inner momentum from mixed
  terms. Furthermore, we introduce the redundant operation $(1-t_p^0)$, viz.
  \setcounter{equation}{0}
  \begin{gather}
    \label{eq:A1}
    \widetilde{\Sigma}(p) :=
      \frac{6(2\pi )^8}{g^2}\Sigma (p)  =  \int\!\!\d^4q\; \left( 1-t_p^2\right)
    \int_0^1\!\!\!\d z \int\!\!\d^4k\; \left( 1-t_p^0\right) \nonumber \\
      \frac{1}{\{ (1-z)[(p-k-q)^2-m^2]+z(k^2-m^2)\}^2} \;
    \frac{1}{q^2-m^2} \nonumber \\
      = \int\!\!\d^4q\; \left( 1-t_p^2\right)
    \int_0^1\!\!\!\d z \int\!\!\d^4\tilde{k}\; \left( 1-t_p^0\right) 
      \frac{1}{\{ \tilde{k}^2+z(1-z)(p-q)^2-m^2\}^2}\; \frac{1}{q^2-m^2}
      \nonumber \\
    = 2\I \pi^2 \int\!\!\d^4q\; \left( 1-t_p^2\right)\int_0^1\!\!\!\d z
    \int_0^\infty\!\!\! 
    \d\rho\; \left( 1-t_p^0\right) \frac{\rho^3}{\{\rho^2-z(1-z)(p-q)^2+m^2\}^2}
    \frac{1}{q^2-m^2} \nonumber \\
    = \I\pi^2 \int\!\!\d^4q\; \left( 1-t_p^2\right)\int_0^1\!\!\!\d z \;
      \ln\left(\frac{m^2-z(1-z)q^2}{m^2-z(1-z)(p-q)^2}\right)\; \frac{1}{q^2-m^2}
  \end{gather}
  Doing a partial integration in the integral over the parameter $z$, and
  introducing the abbreviation $\bar{m}^2:=m^2/(z(1-z))$, yields successively
  \begin{gather*}
    \widetilde{\Sigma}(p) = -\I\pi^2  \int\!\!\d^4q\; \left(
      1-t_p^2\right)\int_0^1\!\!\!\d z\; 
    \frac{z(1-2z)m^2(p^2-2pq)}{[m^2-z(1-z)q^2][m^2-z(1-z)(p-q)^2][q^2-m^2]}\\
    =  -\I\pi^2  \int\!\!\d^4q\; \left( 1-t_p^2\right)\int_0^1\!\!\!\d z\; 
    \frac{zm^2}{z^2(1-z)^2}\;
    \frac{(1-2z)(p^2-2pq)}{[q^2-\bar{m}^2][(p-q)^2-\bar{m}^2][q^2-m^2]}
  \end{gather*}
  In evaluating the integration over the momentum~$q$ one introduces two more
  Feynman parameters $x$ and $y$ such as to obtain
  \begin{gather*}
    \widetilde{\Sigma}(p) = -2\I\pi^2\int_0^1\!\!\!\d z\int_0^1\!\!\!\d x
    \int_0^{1-x}\!\!\!\d y \int\!\!\d^4q\; \left( 1-t_p^2\right)
    \frac{m^2}{z(1-z)^2}  \\
    \frac{(1-2z)(p^2-2pq)}{\{ (1-x-y)(q^2-\bar{m}^2) + x[(p-q)^2-\bar{m}^2]
      +y(q^2-m^2)\}^3} \; .
  \end{gather*}
  Translation of the integration variable $q\mapsto q+xp$ decouples the
  remaining internal momentum from the external momentum~$p$ so that one obtains
  \begin{gather*}
    \widetilde{\Sigma}(p) = -2\I\pi^2\int_0^1\!\!\!\d z\int_0^1\!\!\!\d x
    \int_0^{1-x}\!\!\!\d y \int\!\!\d^4q\; \left( 1-t_p^2\right) \\
    \frac{m^2}{z(1-z)^2} \frac{(1-2z)(p^2-2pq-2xp^2)}{\{
      q^2-x^2p^2-(1-y)\bar{m}^2-ym^2+xp^2\}^3} \\
    =(-2\I\pi^2)^2 \int_0^1\!\!\!\d z\int_0^1\!\!\!\d x
    \int_0^{1-x}\!\!\!\d y \int_0^\infty\!\!\!\d r\; r^3
    \frac{p^2m^2}{z(1-z)^2}\\
    \left( 1-t_p^0\right) 
      \frac{(1-2z)(1-2x)}{r^2+x^2p^2+(1-y)\bar{m}^2+ym^2-xp^2\}^3} \\
     = \pi^4 \int_0^1\!\!\!\d z \int_0^1\!\!\!\d x \int_0^{1-x}\!\!\!\d y \;
    \frac{m^2p^2}{z(1-z)^2} \\
    \frac{(1-2z)(1-2x)x(x-1)p^2}
    {[x(x-1)p^2+(1-y)\bar{m}^2+ym^2][(1-y)\bar{m}^2+ym^2]} \\
     = \pi^4 \int_0^1\!\!\!\d z \int_0^1\!\!\!\d x \int_0^{1-x}\!\!\!\d y \;
       \frac{1}{(1-y)m^2+z(1-z)ym^2} \\
    \frac{z(1-2z)(1-2x)x(x-1)m^2p^4}{x(x-1)z(1-z)p^2+(1-y)m^2+z(1-z)ym^2}    
  \end{gather*}
  The integration over the parameter $y$ yields
  \begin{gather*}
    \widetilde{\Sigma}(p) = \pi^4 \int_0^1\!\!\!\d z\int_0^1\!\!\!\d x \;
    \frac{(1-2z)(1-2x)p^2}{(z-1)(1-z+z^2)} \\
    \ln\left(\frac{\left( -x-(1-x)z(1-z)\right)\left( x(1-x)z(1-z)p^2-m^2\right)}
      {-xz(1-x)(1-z)p^2+m^2-(1-x)(1-z+z^2)m^2} \right) \; .
  \end{gather*}
  Finally, one notices that the following integral vanishes, by the antisymmetry
  of the integrand under $x\longleftrightarrow (1-x)$,
  \begin{equation*}
    \int_0^1\!\!\!\d z\int_0^1\!\!\!\d x \;
    \frac{(1-2z)(1-2x)p^2}{(z-1)(1-z+z^2)} \ln\left( \frac{x(1-x)z(1-z)p^2-m^2}
      {-m^2} \right) = 0 \; .
  \end{equation*}
  Making use of this fact one obtains the result
  \begin{gather}
    \widetilde{\Sigma}(p) \equiv \frac{6(2\pi )^8}{g^2}\Sigma (p)  =
    \pi^4  \int_0^1\!\!\!\d z\int_0^1\!\!\!\d x \;
    \frac{(1-2z)(1-2x)p^2}{(z-1)(1-z+z^2)} \nonumber \\
    \ln\left( \frac{\left[ x+(1-x)z(1-z)\right] m^2}
      {-x(1-x)z(1-z)p^2+(1-x)z(1-z)m^2+xm^2} \right) \; .
  \end{gather}
  This is the result shown in~\Gl{23b}.

  \end{document}